\documentclass[12pt]{article}
\usepackage{graphicx}

\begin{document}

\title{Spontaneous Emergence of Spatial Patterns in a Predator-Prey Model}

\author{M. V. Carneiro \footnote{marcus@stout.ufla.br} and I. C. Charret \footnote{iraziet@stout.ufla.br}\\
Departamento de Ci\^encias Exatas \\ Universidade Federal de Lavras \\ P. O. Box 3707, Zip Code 37200-000, Lavras-MG,  Brazil.} 

\maketitle
\begin{abstract}
We present studies for an individual based model of three interacting populations whose individuals are 
mobile in a 2D-lattice. We focus on the pattern formation in the spatial distributions of the populations. 
Also relevant is the relationship between pattern formation and features of the populations' time series. 
Our model displays travelling waves solutions, clustering and uniform distributions, all related to 
the parameters values. We also observed that the regeneration rate, the parameter associated to the primary 
level of trophic chain, the plants, regulated the presence of predators, as well as the type of spatial 
configuration.

\end{abstract}

\section {Introduction}

Mathematical modelling of population dynamics is widely recognized as a useful tool in the investigation 
of many interesting features found in the organization of individuals in nature \cite{lotka}. 
In order to study the distribution of individuals in their habitat, it is essential to take into account 
factors such as individual's mobility and hunting and escaping skills.

When populations are approximately treated as continuous functions of space and time, 
individual's motion appear in the equations that governs populations distributions as 
diffusion terms. Diffusion is a very common natural phenomena in many areas of science \cite{diversos}.
Thus, much can be learned, by analogy, about the distribution of individuals in their habitats
from other diffusive phenomena. For instance, if interacting populations are described by sets of reaction-diffusion equations, 
it is possible to infer that individuals may be distributed heterogeneously even in homogeneous habitats. 
Other possible phenomena are travelling waves and chaos  \cite{Keitt}.   

Predator-prey models can have their stability properties changed by diffusive terms.
It is been stated by Wilson and De Roos  \cite{Wilson} that spatial predator-prey systems are 
considerably more stable than aspatial ones. 
Originally, population dynamics models used to be formulated in terms of differential equations \cite{lotka}.
This allowed the vast set of analytical methods developed to treat problems in many other 
areas of science and engineering to be applied to ecology. With the advent of cheap computing 
power, it became possible to build more sophisticated models that do not translate easily into 
differential equations or would result in equations too difficult to solve.

Some strategies of analysing and simulating these models includes Individual 
Based Models (IBM) using cellular automata \cite{sherratt}; Individual Based Models 
\cite{Wilson}  without cellular automata; applying mean 
fields approximation \cite{Aguiar}. 
Finite element methods and pertubative methods \cite{King} are other alternative approaches to 
understand these systems. Our choice in this work is to use Individual Based 
Models (IBM) with cellular automata. 
 It consists on applying simple rules inspired in natural events of real system on
a discrete group of individuals lying over a discrete finite lattice.
These rules are organized as a set of events and determines how individuals will behave in each
event such as reproduction and hunting events. We are interested to investigate the global
response of the system.

The main advantage of IBM is the possibility of accounting for many additional features observed on real systems without increasing the computational cost exponentially, such as time 
delayed effects, history-dependent models \cite{Gerami} and attributing to each individual a particular information, like genetics and age \cite{thadeu,meu}.

Our work focus mainly on the spatial patterns that emerge in an open three-trophic food chain 
and their relationship to the population time series. Keitt et al. \cite{keitt2} discussed emergent patterns in 
diffusion-limited predator-prey interaction introducing spatial heterogeneity 
in the model. We observed spatial patterns without this mechanism. Our model presents self 
organization derived mainly from the dynamics to the system. 
We propose an IBM consisting of a fixed plant population, 
a herbivore population, which feeds from the plant population and is able
to diffuse through the system, and a predator population which feeds on herbivores 
in order to reproduce and is able to diffuse through the system as well.

This paper is organized as follows. In the next section, we present a  
description of the model for a three trophic predator-prey system that inspired this work. 
The simulation method is presented and the details of implementation of the cellular 
automata rules are described. In section 3, it is presented results and discussion about the main points of the 
paper and in section 4 we presented our conclusions and future perspectives.

\section{The Model}

Based on the simplest actions of individuals on nature, we proposed the following rules for the 
cellular automata:

\begin{itemize}

\item {\it Movement:} Predators and herbivores can have different probabilities to 
move to a neighbouring site. 
In the simulation, each individual 
receive a random number to decide its next location on the time $t+1$. There are five possibilities related 
to the diffusive rates $d_{1}$ and $d_{2}$ : stay at 
the original site, or going to up, down, left or right neighbours. 
Initially, there is a probability of leaving the original cell wich is divided by the four first neighbours.
The complement of this probability is the chance of being at the same cell.
If two individuals migrate to the same 
neighbour empty cell, only  one individual will remain on the site due to the carrying capacity, 
considered equal to one individual per site. 
If all neighborhood is already full, the individual is forced to stay at its 
original position. As mentioned above, the cell can be occupied by a herbivore, by a predator or both.

\item {\it Natural Death:} 

It is done a draw to each individual of dying with a probability constant set in the simulation. 
Selected individuals are removed from the system on the next time step. 
Plants population does not have natural death draws, however, there is a 
mechanism in its growth rules that prevent it grow exponentially.

\item {\it Plants Growth:} The rule of plants reproduction are quiet simple. Plants has a constant growth 
rate and a carrying capacity. Each site has a float counter that indicates the quantity of resources on 
each time step and it generally changes after a herbivore visit. All sites in the lattice has its plants 
counter incresead by the fixed constant value determined in the program without no draws until it reaches 
the limit imposed by the carrying capacity. The plant growth in a current site does not affect the neighbour 
cells.

\item {\it Plants Gathering:} 

When the herbivore comes to a site, the 
main rule is to gather the maximum quantity of food until hunger counter is reduced to zero. If the hunger 
is less than the resources quantity, it eats what it needs, setting its counter to zero and leaving the 
remaing food quantity in the site. If the opposite is verified, it eats all the site resources setting the 
site's counter to zero and keep with the hunger counter set to the difference between the two quantities. 
The plants gathering process does not have any draws.

\item {\it Predation:} Predation occurs when herbivore and predator share the same cell. In this
cases, it is done a draw to decide if the hunter is succesfull. In positive cases,
predators have the hunger counter set to zero and prey will leave the system in the next time step. 
In negative cases, nothing happens.

\item {\it Reproduction:} After all events described above has been applied, the populations are allowed 
to reproduce. Herbivores and predators have similar reproduction rules. Both populations have different 
probability to reproduce and only individuals whose hunger counter is zero will have chance to do it. 
As we did in the movimentation rule, if two herbivores or two predators reproduce in the same 
empty cell, only one child will stay in the site. Predators and herbivores can reproduce one child that 
will be placed in one of the available neighbour sites, like the movimentation rule. If all neighbour 
sites are already occupied the birth will be canceled.

\end{itemize}

Time will be discretized into time steps that could be interpreted as generations of the population. 
Space will be discretized into a homogenous 2-D lattice formed by cells. The cell can be empty, 
occupied by a prey, occupied by a predator or occupied by both. To each cell is associated a real number 
that represents the quantity of plant resources available. The sum of the values of all cells represents the 
total of the plant population. Each mobile individual has a intrinsic counter called hunger which is 
incremented every time step at this event. The bigger this value becomes, the longer is the time that 
individual has not fed. The neighbourhood adopted is the Von Neumamm type \cite{neumann,birgitt}, 
that includes four first neighbours. The boundary conditions proposed are periodic to reduce finite size 
effects \cite{keitt2}. Events will be applied to each individual on the each time step in the simulation.
In each event, there is a draw using an probability parameter. 
The rates of each term will be worked as probabilities and the event applied to individuals are stochastic. 
For example, on the natural death event, it will be make a draw to each individual to die with a probability
assigned to the simulation. 
A similar procedure has been proposed by Wilson and de 
Ross \cite{Wilson}. They proposed that in an event, it must be done a draw to each individual, however due 
to the limited computational resources they have adopted another similar method for the events drawings.

The individuals are initially distributed over around the lattice. 
Events are applied in the order they are described above but the order that events were applyied is irrelevant.
The updating of the populations occurs at the end of each time step. Individuals which died on the time $t$
will be removed from the system,  new individuals will be placed on the region and the 
position of the existing individuals will be updated. Individuals that were select to die in a given time step still reproduce at that step, being removed from the population at the end of each round.

The results were obtained in regular square lattices with $100 \times 100$ cells, with periodic boundary conditions. Each simulation consisted of $10^{6}$ time steps spending about $20$ minutes in machines with the following characteristics: 32-bits Athlon MP processor with 512 MB using GNU C++ compiler 4.0.

\section{Results and Discussion}

We have tested several cases with different sets of probability values to find all possible steady 
states of the system: coexistence of three populations, extinction of predators and extinction of 
herbivores and predators. Our analisys show that variation of the plant regeneration rate is enough to sweep through all possible steady states. As the mobile populations relies on the 
plant population, this constant can control directly the presence of any population on the system.
Following results are organized according to this parameter. Discussions about the effects of this parameter
will be presented from small values of regeneration rate, where both populations go to extinction, to higher values, where we have the coexistence among three species. We are concerned to focus on states whose behaviour are close to the critical states. We mean by critical state a point in the phase space which some small changes of parameters values can lead to a different steady state, such as extinction of one population or distinct 
behaviour of the time series.

One important consequence of discretization of individuals is the increase of the possibility of extinction 
of the populations. It can be observed on the numerical solution of the continuum model, that populations could 
assume very small values tending to zero, however in the following instants the population value increases and 
keeps oscillating without going to extinction. On discrete individual simulation, this can not be verified 
because the populations are not capable to increase their value after they have gone to extinction. Simulation 
tests are very sensible to stochasticity when the populations are small.

\begin{figure}[ht]
\centerline{\rotatebox{-90}{\includegraphics[scale=0.30]{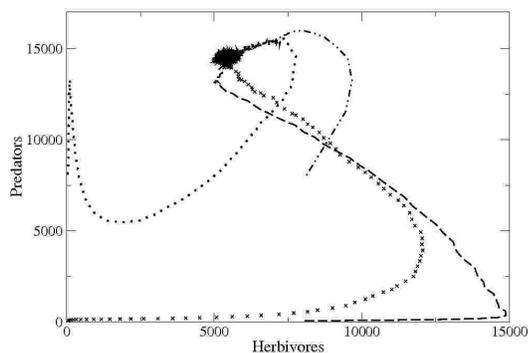}}}
\caption{\small Phase space portrait of plants and herbivores populations of four simulations with parameter values 
given in table \ref{table1}, set C1. Values of initial populations are differents in all simulations.}
\label{popinicial}
\end{figure}

Figure \ref{popinicial} shows the phase portrait of four simulations using the same set of parameter values with different initial populations. All simulations predicts the same steady point, independently of the initial conditions. This result indicates that the system have a wide basin of atraction. However, herbivores and predators populations have a higher probabilities to go to extinction when they start from small values. Extinction cases frequently occurs when populations values are aproximately equal to the stocasticity fluctuations.

\begin{figure}
\centerline{\rotatebox{-90}{\includegraphics[scale=0.30]{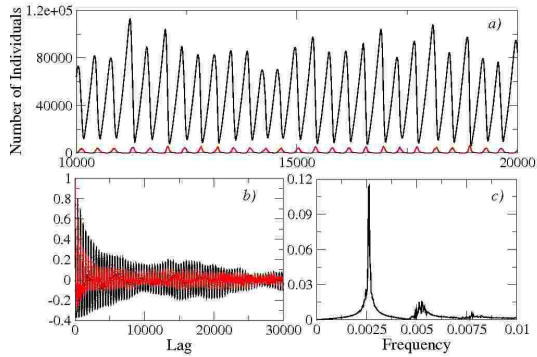}}}
\caption{{\small Results of simulation using the set of parameters C1 in table \ref{table1} and regeneration rate of 0.02. Predators go to extinction at begginning of the simulation. In (a) is shown the time series; in (b) is shown the autocorrelation function; in (c) is shown the frequency spectrum of plants population. Herbibores population has the same spectrum.Black curves represent plants population and gray curves represent herbivores populations. Simulation has taken $10^6$ time steps.}}
\label{partida}
\end{figure}

\begin{table}
\begin{center}
\begin{tabular}{  l  |  c  |  c  |  c  |  c  }
\hline
Skills & \hspace{0.2cm} C1 \hspace{0.2cm} & \hspace{0.2cm} C2 \hspace{0.2cm} &  \hspace{0.2cm} C3 \hspace{0.2cm} &  \hspace{0.2cm} C4  \hspace{0.2cm}\\ 
\hline
Hunter & 40 & 40 & 90 & 60  \\ 
Prey reproduction & 80 & 80 & 50 & 80 \\
Predator reproduction \hspace{0.3cm} & 40 & 40 & 85 & 50 \\
Prey death & 1 & 1 & 5 & 5 \\
Predator death & 5 & 5 & 5 & 5 \\
Prey mobility & 80 & 40 & 80 & 80 \\
Predator mobility & 80 & 80 & 80 & 80 \\
Carrying capacity & 20 & 20 & 20 & 20 \\
\hline
\end{tabular}
\caption{\small Sets of probabilities for the parameters used in the simulations.}
\label{table1}
\end{center}
\end{table}

\begin{figure}
\centerline{{\includegraphics[scale=0.26]{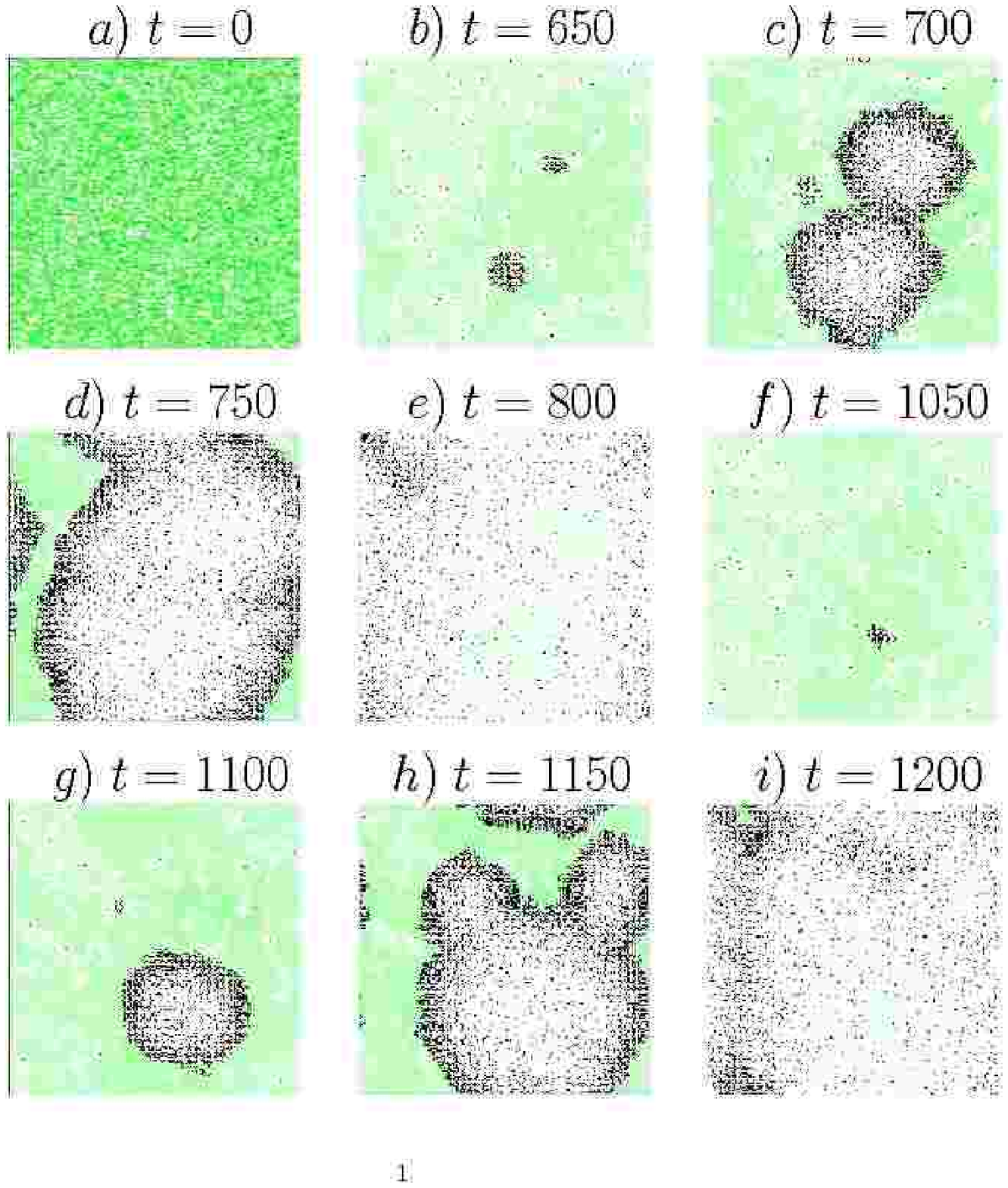}}}
\caption{{\small  Spatial distribution of simulations adopting a regeneration rate $0.02$ and the set of 
parameters C1 in table \ref{table1}. Each figure corresponds to a different time-step of the same simulation.
The main feature is the appearance of big herbivore waves throughout the lattice. Plant population is 
represented in gray scale and herbivore in black.}}
\label{distone}
\end{figure}

We observed two non trivial steady states. The first one is the  coexistence of plants and 
herbivores populations. The other is the coexistence of the three species. Changing the 
regeneration rate of the plant population and keeping other probabilities constant is possible to verify 
some distinct population behaviour and to control the existence of herbivore and predator population on the region.

\subsection{Coexistence of two populations}

\begin{figure}
\centerline{\rotatebox{-90}{\includegraphics[scale=0.30]{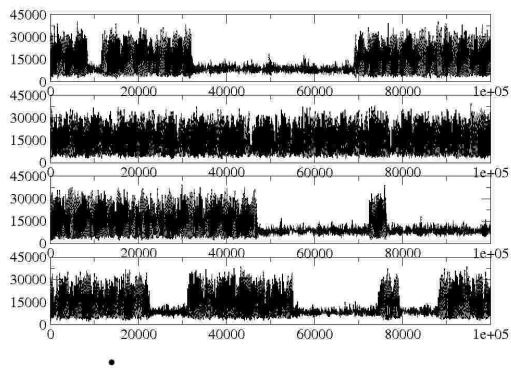}}}
\caption{{\small Results of four simulations adopting the set of parameters C1 in table \ref{table1} and 
the same initial conditions with the regeneration rate $0.033$. 
The occurrance of the phase transition is distinct in each one of the simulations.}}
\label{transgraphs}
\end{figure}

For the parameter set C1 shown in table \ref{table1} and regeneration rate below $0.0175$, both 
mobile populations go to extinction and plants grows logistically.

For regeneration rates between $0.0175$ 
and $0.3$, prey are capable to survive and predators still go to extinction. This fact shows the existence 
of multiple steady states of this system.

\begin{figure}
\centerline{\rotatebox{-90}{\includegraphics[scale=0.30]{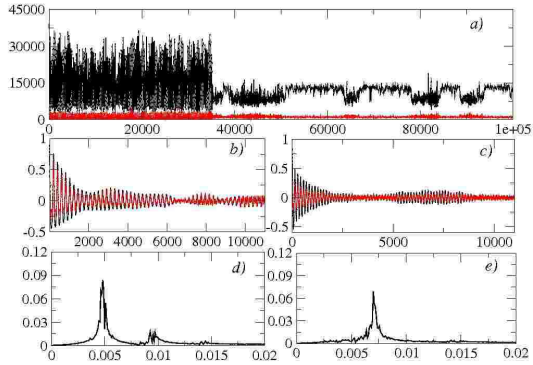}}}
\caption{{\small Frequency analysis of the two distinct parts of the time series of populations obtained with 
parameters set C1 and the regeneration rate $0.033$. In figure (a) is shown the time series with two distincts regions. First part corresponds to a region with high amplitude and second corresponds to a region with low amplitude.
Figure (b) and (c) shows the autocorrelation function for high and low amplitude intervals respectivaly
Figure (d) shows the presence of one characteristic frequency and two harmonics of it and is related to first part. Figure (e) presents only one frequency, which is 
different from the previous frenquencies found at the first part.}}
\label{analysis}
\end{figure}

\begin{figure}
\centerline{{\includegraphics[scale=0.35]{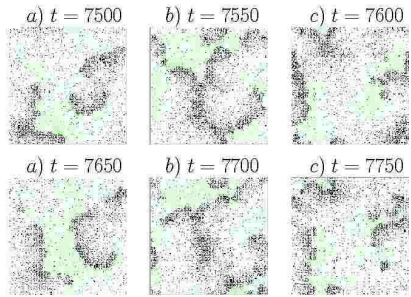}}}
\caption{{\small  Spatial distribution of simulations adopting a regeneration rate $0.033$ and the set of 
parameters C1 in table \ref{table1}. Each figure corresponds to a different time-step of the same simulation.
Figure shows a interval that system oscillates with a low amplitude. Populational waves is smaller than
the case with high amplitude oscillation and indivuals are more distributed through the region.}}
\label{table2}
\end{figure}

Figure \ref{partida} shows the results of the simulation adopting the parameters values C1 in table \ref{table1} and the regeneration rate $0.02$. We can observe that plants and herbivores populations time series in part $(a)$ show an oscillatory behaviour. The fourier transform of the 
autocorrelation function in part $(c)$ clearly indicates the presence of one fundamental frequency for plant population and two harmonics higher. This fact indicates the existence of only one characteristic time in 
this system. This behaviour is the same for both populations. Some simulations around this value of regeneration rate show only one harmonic for the same frequency.

\begin{figure}
\centerline{\rotatebox{-90}{\includegraphics[scale=0.30]{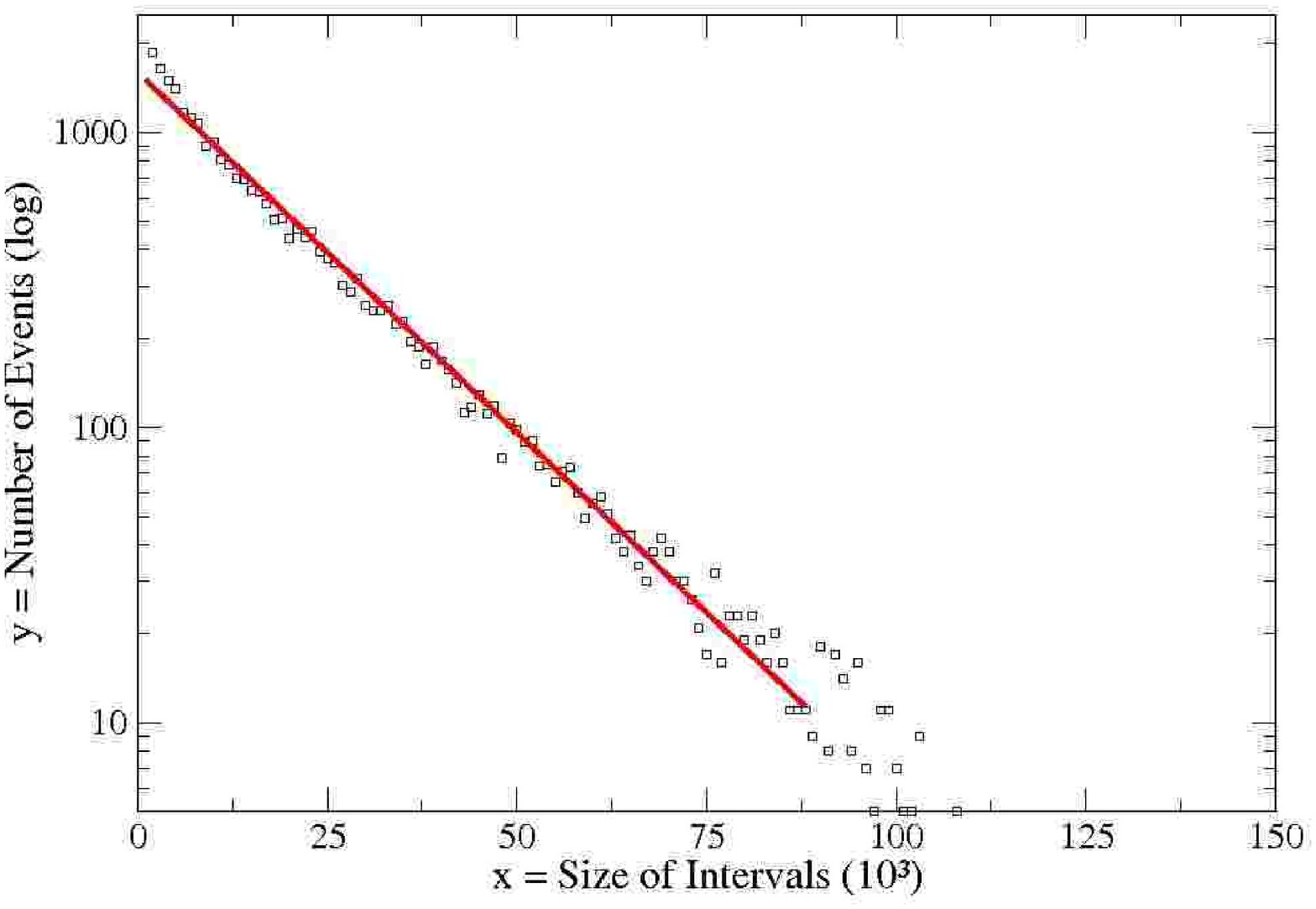}}}
\caption{{\small Distribution of the size of intervals. The exponential curve adjusted from the data is  $y = 1596e^{-0.00562x}$}}
\label{histograma}
\end{figure}

\begin{figure}
\centerline{\rotatebox{-90}{\includegraphics[scale=0.30]{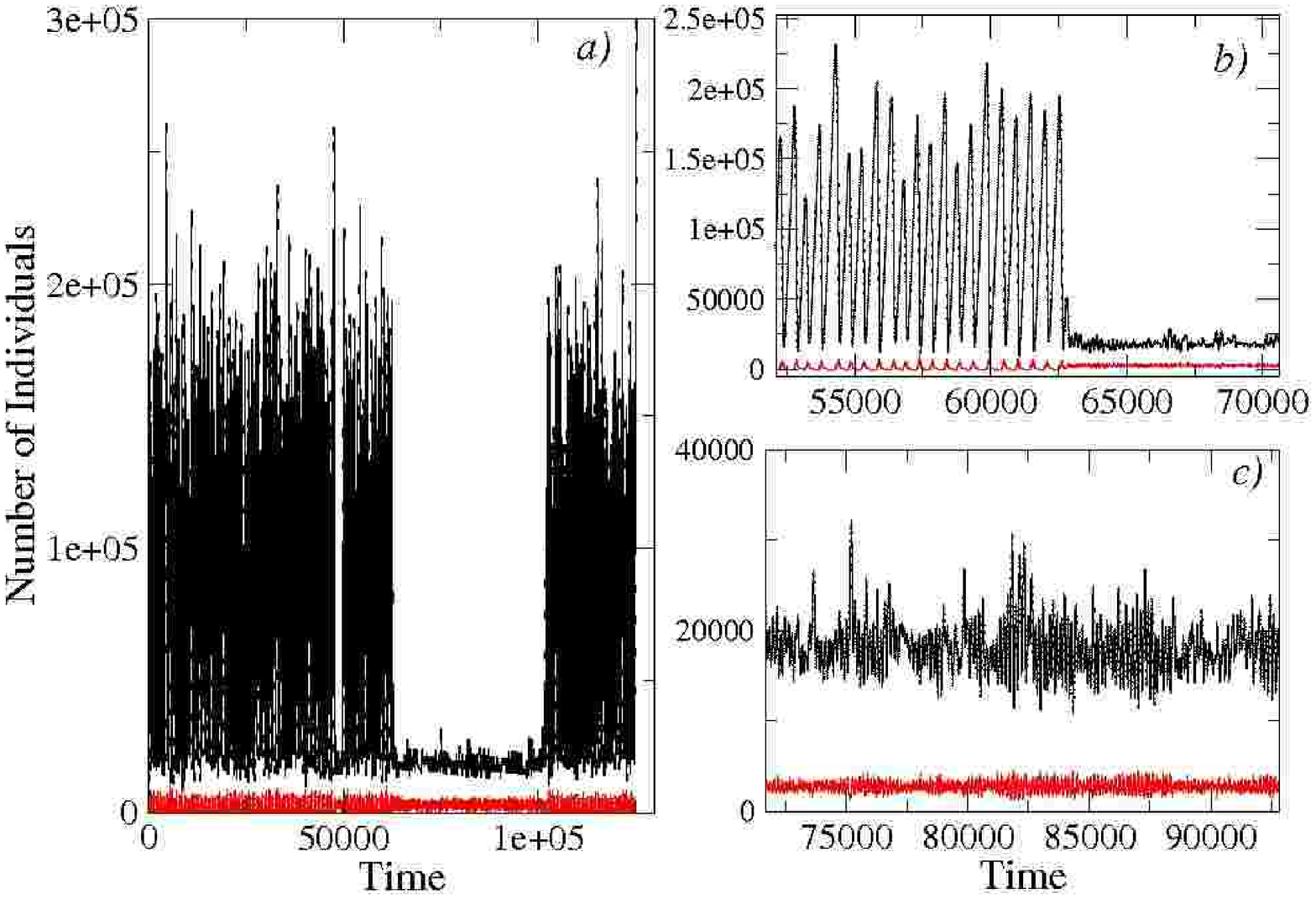}}}
\caption{{\small Simulation with different parameter shows the same transition effect. 
Graphs (b) and (c) shows a "zoom" in different parts of the graph (a). After the time 125000, 
herbivore population goes to extiction. Results obtained with the parameter set C2 in table \ref{table1} and 
regeneration rate of $0.025$.}}
\label{serietrans}
\end{figure}

Comparing time series and spatial distribution, we observe the emergence of population waves migrating 
towards the food gradient as we can see in figure \ref{distone}. These populational waves are non-linear and one of the effects of non-linearity in this case is the individuals annhilation in the wavefronts. Waves collisions are very commom due to the periodic boundary conditions. The behaviour of the times series in  figure \ref{partida} corroborates our observations of the individuals spatial distribution in figure \ref{distone}. The 
populational time series has a huge oscillatory amplitude that is verified on spatial distribution. Small 
populational clusters emerge and diffuse on the habitat like a wave, leaving some individuals behind on the path.
As the wave travels through space, the number of individuals grows up fast until it collides to
itself . With the collision, many individuals die due to the carrying 
capacity of the cells and the local lack of food. Population decrease to small numbers until the plants 
resources regenerate. The probability of new waves appear is linked to regeneration rate
of plants: as it is increased more waves start to appear in different places. 
Oscillatory behaviour is verified until the regeneration rate reaches values around $0.034$.

\begin{figure}
\centerline{\rotatebox{-90}{\includegraphics[scale=0.30]{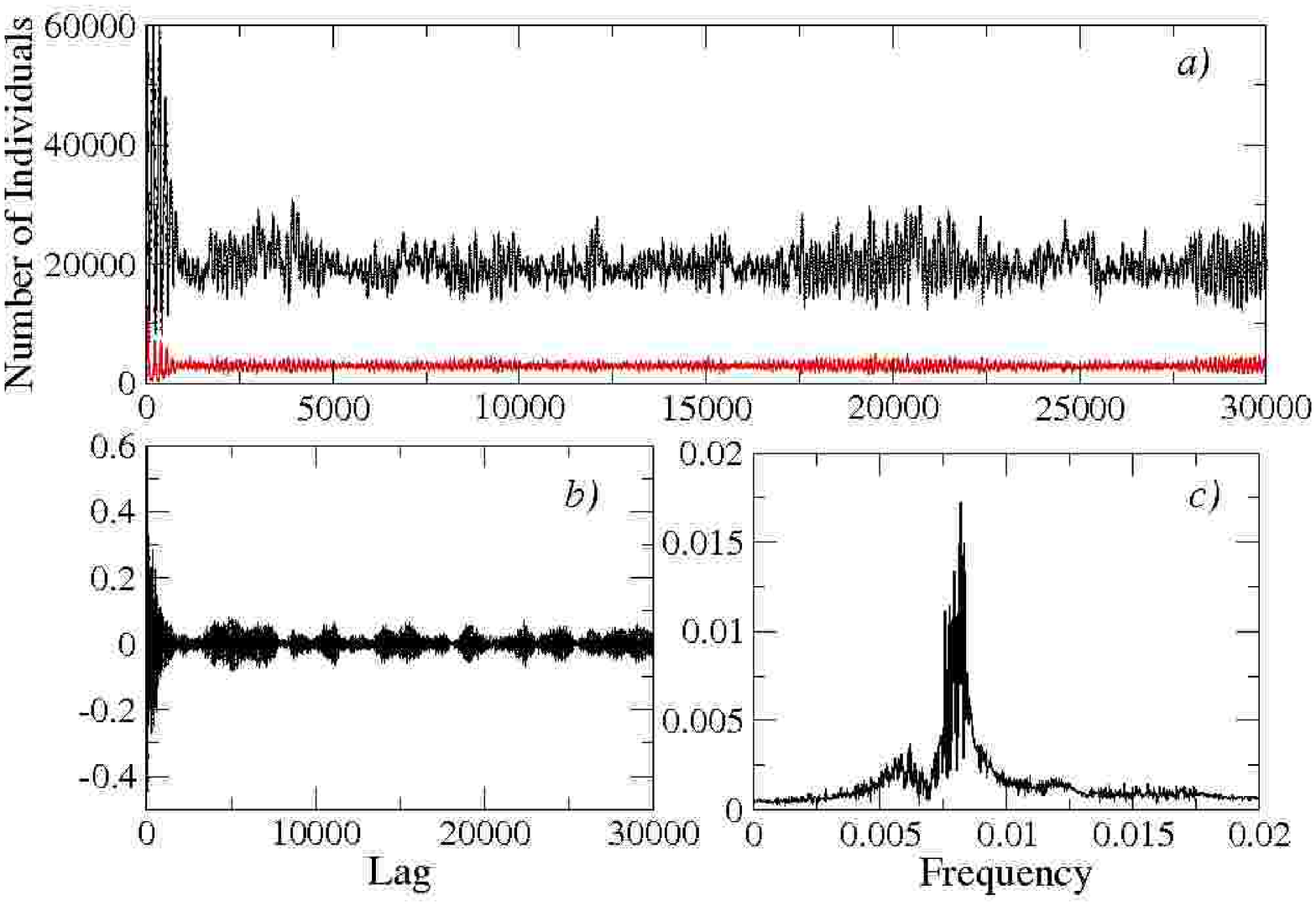}}}
\caption{{\small Results obtained with set C1 and a regeneration rate above $0.05$. It shows 
a more stable behaviour. Graph (a) shows the time series, (b) the autocorrelation function, (c) the fourier transform. As we can see at the fourier transform,
the oscillation of the time series does not show a relevant frequency.}}
\label{stabler}
\end{figure}

In the region of the regeneration rate between $0.03$ and $0.04$, we observe clearly in the figure 
\ref{transgraphs} a change of behaviour in the time series. Figure \ref{transgraphs} shows a few plant time series for a regeneration rate of $0.033$. Plant population changes its oscillation from a 
high amplitude to a low amplitude in an uncertain time. The oscillatory behaviour with low amplitude seems to be more stable and prey populations do not go to extinction in any time.
Figure \ref{analysis} shows an example of this change with a respectively frequency analysis. The high amplitude region of the time series presents one characteristic frequency and one harmonic. The low amplitude region of the time series presents only one frequency. This frequency is different from the frequency found in the first part. Populational waves on the spatial distribution are still 
verified after the transition time, in the region with the low amplitude and is shown on figure \ref{table2}. However, the herbivore population is much less localized on the space and the waves are not so big 
than in the previous cases, with low regeneration rate shown in figure \ref{distone}.

Like the ones show in figure \ref{transgraphs}, we ran 3500 simulations to identify features of the 
transition to the occurrance of this phenomena. We measured the size of each low amplitude oscillation 
regions larger than $1000$ time steps. The results is shown in figure \ref{histograma}. The distribution of the occurrance of the size intervals obeys an exponential function so that  small intervals occurs more frequently than bigger ones, showing that the system oscillate between the two states, high and low amplitude, nondeterministically. There is no way to predict when this phenomena of transition is going to happen.

\begin{figure}
\centerline{\rotatebox{-90}{\includegraphics[scale=0.30]{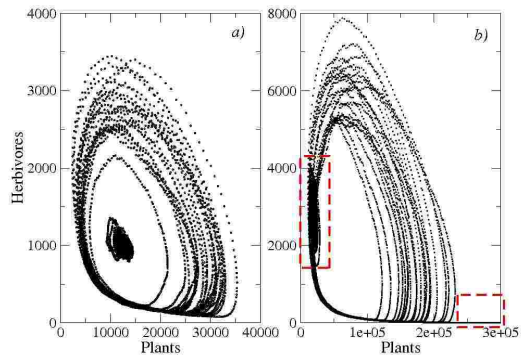}}}
\caption{{\small Two phase portrait from distinct probabilities set. First one is the result from simulation
presented by the figure \ref{analysis}, and the second one is the result from the simulation of the figure \ref{serietrans}.
The small rectangles in (b) indicates the points where the simulation stays during some time steps.}}
\label{phaseportrait}
\end{figure}

This behaviour also can be found in other sets of probabilities, as we can see 
in figure \ref{serietrans}. When a population changes its behaviour, the other one also follows it. 
However it is not possible to state which population start this change.

\begin{figure}
\centerline{\rotatebox{-90}{\includegraphics[scale=0.30]{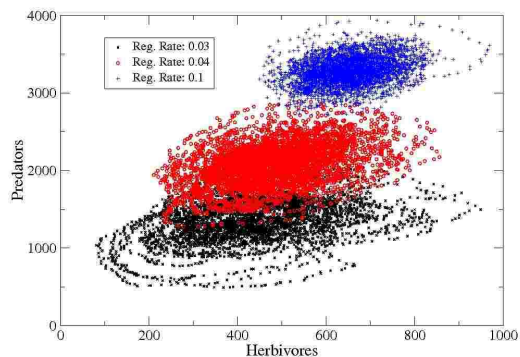}}}
\caption{{\small Phase space portrait of herbivores and predators populations of three simulations with set of parameters C3 in table \ref{table1} and differents values of regeneration rates.}}
\label{espacofase}
\end{figure}

Actually, in this interval comprehending values of regeneration rate between $0.03$ and $0.04$, the time series of the system show behaviours that alternate between the solutions 
found with the regeneration rate under $0.03$, which has a high amplitude oscillations, and the 
solution found  with regeneration rate above $0.04$, that present low amplitude oscillations and more 
stable state. Figure \ref{stabler} shows the same simulation adopting a regeneration rate of $0.05$. This figure corroborate our conclusions about this behaviour.

Figures \ref{phaseportrait} shows a clean phase space portrait corresponding to the 
simulations shown in the figure \ref{analysis} and figure \ref{serietrans}.
We removed the transitory points of time series and displayied the "stationary values" only. 
In the first phase portrait, the system initially oscillates through a closed orbit
around the equilibrium point, characterizing an oscillatory behaviour. Due to stochasticity, it is not 
possible to define clearly the orbit, however it is possible to identify a region where it occurs.
However, it suddenlly jumps to an equilibrium point, that characterize non-oscillatory behaviour for long time intervals. 
Second phase portrait shows a simulation that initially oscillates
through a large orbit and jumps to a point indicated in left rectangle in the figure. Subsequently it returns to the 
larger orbit and goes to extinction after some time steps later, as indicated on the right rectangle in the 
figure. This indicates the system has a multiple steady states and the stochasticity is responsible 
for the transitions between one to another. As we said before, populations have a higher probability to go 
to extinction when it has this high oscillatory behaviour because it assumes small values and stochasticity 
is very relevant.

\subsection{Coexistence of three species}

\begin{figure}
\centerline{\rotatebox{-90}{\includegraphics[scale=0.30]{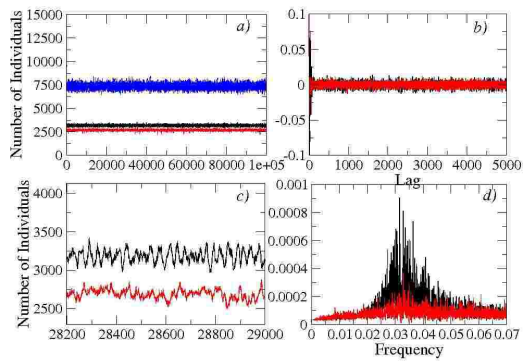}}}
\caption{{\small Results of simulation using C4 set parameters values in table \ref{table1} and regeneration rate $0.2$. In (a) is shown time series for three populations. Plants populations are dark gray, predators are black and herbivores are light gray curve. In (c) is shown a zoom on the time series for predators and herbivores, emphasizing the oscillatory behaviour of both; (d) show the frequency spectrum to the autocorrelation functions (b).}}
\label{uniforme}
\end{figure}

\begin{figure}
\hspace{-2cm}
\centerline{\includegraphics[scale=0.35]{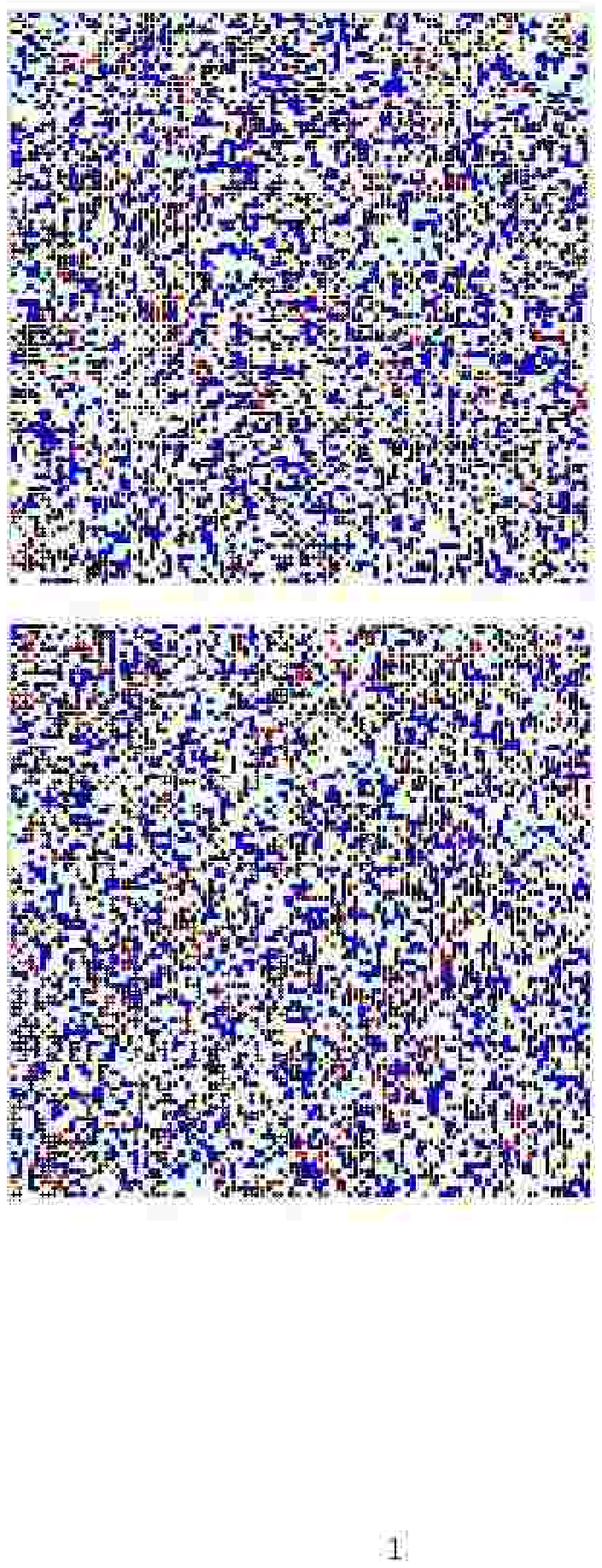}}
\caption{{\small Spatial configurations correspondent to case shown in figure \ref{uniforme}. Plants populations are light gray, predators are black and herbivores are dark gray points.}}
\label{table5}
\end{figure}

As we increase the regeneration rate, the wave behaviour starts to disappear and herbivores spread through the 
lattice in a uniform way. After herbivore population reaches a more stable behaviour, predator are able to 
survive because there is more available food.

\begin{figure}
\centerline{\rotatebox{-90}{\includegraphics[scale=0.30]{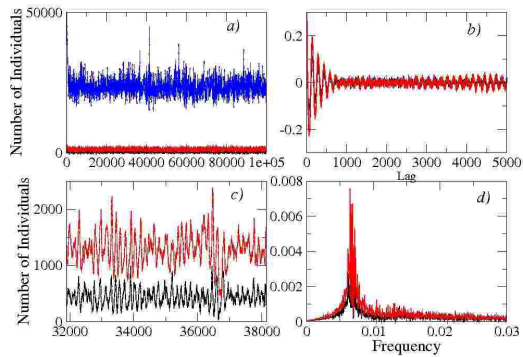}}}
\caption{{\small Results of simulation using C3 set parameters values in table \ref{table1} and regeneration rate $0.03$. In (a) is shown time series for three populations. Plants populations are dark gray, predators are black and herbivores are light gray curve. In (c) is shown a zoom on the time series for predators and herbivores, emphasizing the oscillatory behaviour of both; (d) show the frequency spectrum to the autocorrelation functions (b)}}
\label{aglomerado}
\end{figure}

\begin{figure}
\hspace{-2cm}
\centerline{\includegraphics[scale=0.35]{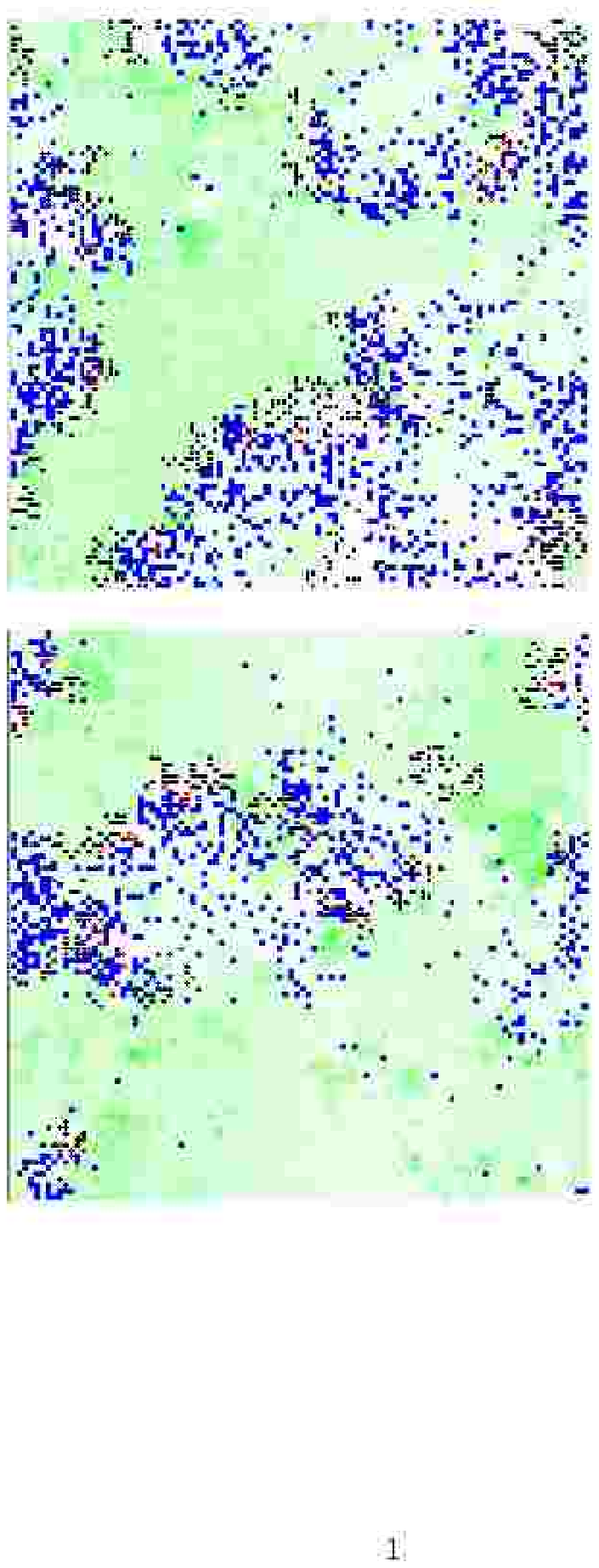}}
\caption{{\small Spatial configurations correspondent to case shown in figure \ref{aglomerado}. Plants populations are light gray, predators are black and herbivores are dark gray points.}}
\label{table4}
\end{figure}
 
In some simulations, we observed that the number of predators are directly related to the regeneration
rate of the plants. As we increase the regeneration rate of the plants we dislocate the equilibrium point 
of the system in a positive way. Figure \ref{espacofase} shows 3 simulations using a distinct set of parameters
adopting different values for the regeneration. We exchange the set of parameters to one which predators 
would survive even though regeneration rate is low. Increasing the regeneration rate also makes the population
reduce the oscillation levels around the equilibrium point leading the system to a more stable state.

We have distinct spatial distribution in the case of coexistence between the three species.
These distributions are directly related with the abundance of each population. 
The constants that affect directly those quantities are regeneration rate, herbivore birth rate,
predator birth rate and hunter rate. The last one is important because according to the
cellular automata rule only predators which ate on the current iteration can reproduce.

Situations with a high herbivore birth rate and a high regeneration rate result in a uniform spatial 
distribution shown on the figure \ref{table5}. Herbivores are constantly reproducing occupying the whole region
and predators are distributed uniformly as well as the herbivores. Time series obtained with this
distribution is shown on figure \ref{uniforme}. We can notice that plant population has a low value due to 
the high number of preys gathering it. Populations has a very stable behavior with
no relevant oscillations. The Fourier transform in the graph $d$ indicates the presence of few 
frequencies with very low weight comparing to ones we obtained previously.
We considered this behaviour as the closest to logistic equation behaviour, since it reaches a point 
in its steady state and it does not have any oscillation neither a characterisct frequency.

However, if we decrease the herbivore availability reducing the birth rate and increase predators
by adding predator birth rate and hunter rate we obtain a different distribution. Herbivore population
is distributed in clumps with predators around it as we can see on figure \ref{table4} . Herbivore population is now rare and predators need to stay near these clumps to reproduce and survive. Our observations
point out that the occurance of clumps is related to the ratio of the population's availability
between prey and predators. Our observations of clumping agree with Kareiva's \cite{kareiva} considerations. He states that one unavoidable outcome of local interactions and local dispersal when we are 
working with cellular automata is clumping. The characteristics of the clumps are strongly related with the 
parameters of the model such as birth and mortality rates.

In these cases, as can we see in figure \ref{aglomerado} plant population reaches higher values due to the small number of the preys and 
predator population has higher values due to favorable parameters set in the simulation. All populations has a higher oscillation and the fourier transform indicates the presence of one
characteristic frequency.

\section{Conclusions}

We studied the spatial behavior of a predator-prey system in a three trophic food chain using 
Individual Based Model. We focused, basically, on the relationship between spatial phenomena and the time
series behavior.

The simulations were run with initial homogenous distributions yielding different kinds of spatial patterns 
spontaneously according to the set of parameters assigned to the simulation.
In all situations treated here, none inhomogenous spatial conditions were applied, so 
all distributions verified are results from the dynamics of the system established by the rules
of cellular automata.

Results allow us to conclude that the regeneration rate of the plants is the critical parameter 
of the system. It is strongly related to the stability of the steady point as well as with the behavior 
of the system at this point.

Adopting low values of regeneration rate and allowing the coexistence between prey and plants, it is 
possible to verify the occurrance of the travelling waves on the spatial distribution.
Time series related to this distribution presents a high oscillation amplitude with a characteristic 
frequency. Increasing this parameter we observe a different kind of travelling waves and distinct time series behavior,
which has a low oscillation amplitude and a different characteristic frequency. The system alternates 
between this two behaviors for some values of regeneration rate. This alternance is nondeterministic
and the distribution of the intervals size of each behavior obeys a exponential function. 
Another possible spatial configuration is the uniform and 
it occurs when resources are abundant and preys grows up reaching the carrying capacity of 
the region.
 
This configuration is more stable and it allows the maintenance of predators. They can not survive in cases
there is herbivore travelling waves.
Results with the coexistence of three species is quite more stable. Situations with
herbivores abundance present uniform spatial distribution, however situations where
predator are smarter and herbivores are rare, present  a formation of populational 
clumps on the region. Unfortunatly, these distinct distribution have similar 
frequency spectrum and none feature could distinguish them.

The problem treated here is completely stochastic and the appearence of pattern in the spatial distributions 
is a consequence of the self organization of the system. In sequence, we are developing a study about the 
association between kind of pattern and the characteristic of the temporal series that describe the system. 
We are trying to produce a tool that can be used to classify the system using experimental information, 
frequently available in a temporal series form.

{\bf Acknowledgements}

It has been partially financial supported by  brazilian agencies FAPEMIG, CNPq and FINEP. The authors wish to thank Dr. A.T. Costa Jr. for his comments throughout this work.

\end{document}